\documentstyle[twocolumn,aps,epsf,prl]{revtex}
\begin{document}
\draft
\preprint{}
%
\twocolumn[\hsize\textwidth\columnwidth\hsize\csname
@twocolumnfalse\endcsname

\title{Structural Stability of Vacancy-ordered Yb$_{2.75}$C$_{60}$}
\author{K. M. Rabe$^1$ and P. H. Citrin$^2$}
\address{$^1$Department of Applied Physics, Yale University, New Haven, CT
06520-8284}
\address{$^2$Bell Laboratories, Lucent Technologies, Murray Hill, NJ 07974}
\date{\today}
\maketitle
\widetext
\begin{abstract}
The fcc-based structure of Yb$_{2.75}$C$_{60}$ is unique among metal-doped
fullerene compounds, exhibiting long-range-ordered vacancies,
significantly off-centered divalent Yb cations, and distorted,
crystallographically inequivalent, orientationally ordered C$_{60}$
anions. A simple electrostatic-energy analysis, which models the
constituents using point charges, is shown to provide insight into how
each of these features stabilizes this unusual crystal structure. The
results have general implications for a variety of other intercalated
metal fullerides.

\end{abstract}

]

\narrowtext

Since 1991, numerous superconducting compounds of C$_{60}$ intercalated
with monovalent alkali and divalent alkaline-earth metals have been
prepared and structurally characterized \cite{review}. In many respects, their
crystal structures can be described by regarding the C$_{60}$ anions as
large, negatively-charged ``atoms" forming simple fcc or bcc lattices,
the interstices of which are filled by metal cations. The stability of
these structures is generally well understood within simple models
dominated by electrostatic energies \cite{jcp,tom}. The analysis of the
structural energetics through more complete first-principles
calculations confirms and refines the simple models \cite{and1}.

Ytterbium metal has also been shown to form a superconducting
compound with C$_{60}$, but with a structure that is considerably more
complex \cite{nature}. It can be viewed as a modification of a hypothetical
Yb$_3$C$_{60}$
structure, analogous to that of K$_3$C$_{60}$ \cite{step} and schematically
illustrated
in Fig. \ref{fig1cap}(a). In this ideal Yb$_3$C$_{60}$ structure, Yb cations
sit at the
centers of the octahedral ($O$) and tetrahedral ($T$) holes created by the
fcc lattice of pure C$_{60}$. Systematically removing 1 of the 8 $T$-cations
leads to the ordered-vacancy structure of Yb$_{2.75}$C$_{60}$ in Fig.
\ref{fig1cap}(b), where
the full unit cell is seen to contain 8 subcells, each with a different
$T$-site vacancy. One such subcell, with the C$_{60}$

\begin{figure}
\epsfxsize=3.0 truein
\caption{(a) Schematic picture of hypothetical fcc Yb$_3$C$_{60}$ structure
(C$_{60}$
anion radius is  3.5 \AA, Yb$^{2+}$ radius is  1.0 \AA). Cations fill all
tetrahedral ($T$) and octahedral ($O$) interstitial sites and are shaded
gray and black, respectively. Removing one $T$-site cation (shown with
arrow) leads to the fcc-based vacancy-ordered Yb$_{2.75}$C$_{60}$ structure in
(b), with the C$_{60}$ anions removed for clarity. Eight subcells comprise
the full unit cell. In the expanded subcell shown, $O$-site cations are
significantly displaced from their ideal, ``centered" positions towards
the nearest vacancy, and each $T$-site cation is displaced less (drawn
exaggerated) along one of the Cartesian axes. Other $O$-site cations have
been displaced towards vacancies in adjacent subcells. (c) Same
Yb$_{2.75}$C$_{60}$ subcell as in (b) but with the Yb cations omitted. The
vacancies create 3 inequivalent types of C$_{60}$ anions, schematically shown
with different shadings and shapes. Each anion rotates about the local axis
indicated, to maximize the number of pentagon faces oriented towards the
surrounding cations.}
\label{fig1cap}
\end{figure}

\begin{figure}
\epsfxsize=3.3 truein
\centerline{\epsfbox{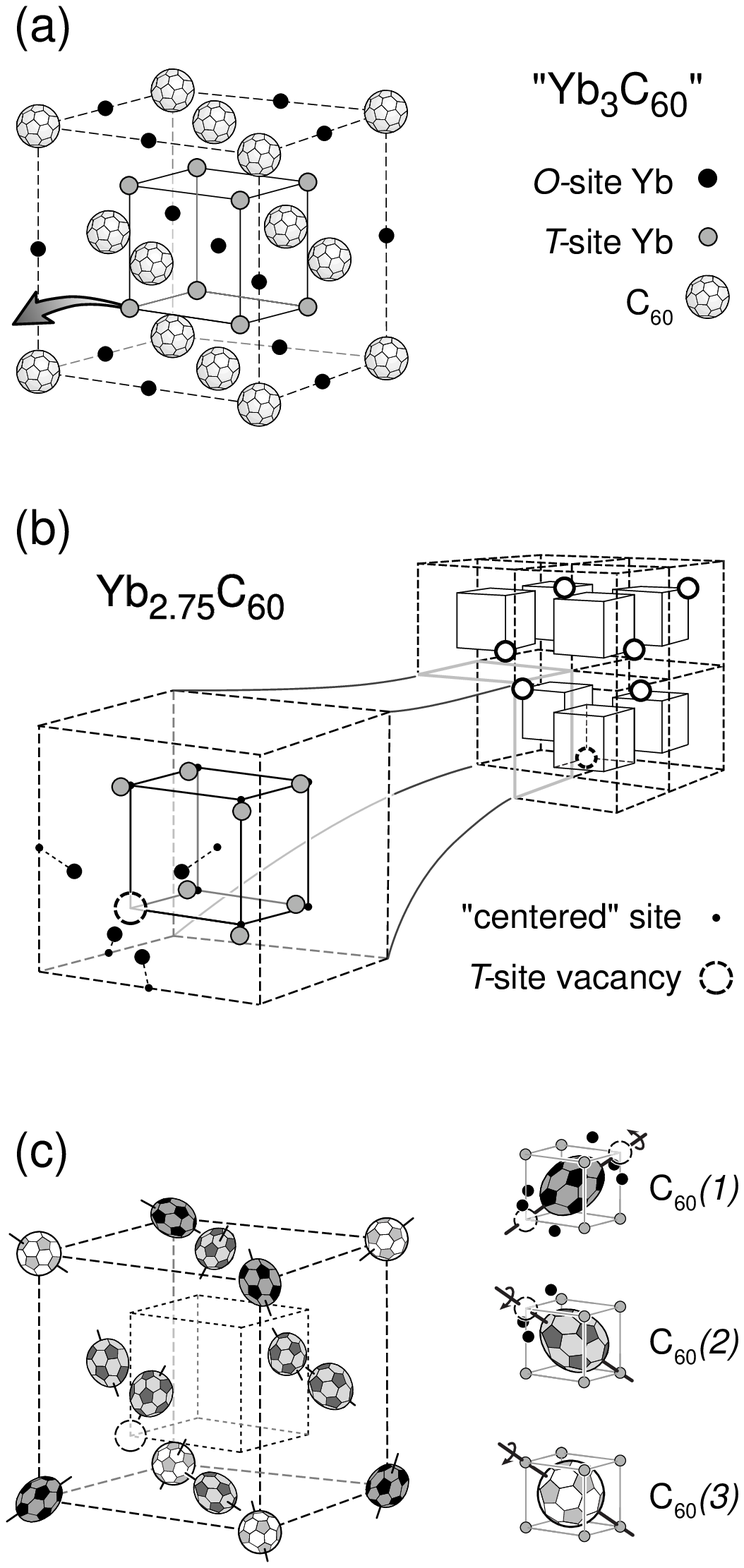}}
\label{fig1}
\end{figure}

\noindent
anions removed for
clarity,
highlights the large displacements of the $O$-cations towards
their nearest-neighbor (NN) vacancy and the smaller displacements of the
$T$-cations along Cartesian axes \cite{nature}. The 3 types of
crystallographically
inequivalent C$_{60}$ anions with 2, 1, or 0 neighboring vacancies, and their
correspondingly different coordinations of 12, 10, and 8 Yb cations, are
shown in Fig \ref{fig1cap}(c). Also included schematically are the different
distortions of the anions \cite{citrin} and the local axes about which they
rotate in order to orient their pentagonal faces towards the NN cations.

In this Letter, we identify the underlying reasons for the stability
of this apparently exotic phase. A simple electrostatic model is used to
evaluate the energies associated with vacancy formation,
cation displacements, and C$_{60}$ orientations and distortions. Our
analysis shows that the latter relaxation energies more than
compensate the cost of creating a Yb vacancy. The formation of
Yb$_{2.75}$C$_{60}$
is therefore seen to fit naturally into a unified understanding of both the
stability of metal-doped C$_{60}$ compounds and the role of particular
structural defects.

Our investigation focuses on understanding why the ideal Yb$_3$C$_{60}$
structure in Fig \ref{fig1cap}(a) is not formed but is instead found to phase
separate into the observed Yb$_{2.75}$C$_{60}$ structure and Yb metal. For
these
two structures, we compare the relative energies of steps in a
Born-Haber cycle, taken with respect to a reference system of undoped
fcc C$_{60}$ and isolated neutral Yb atoms. The steps are (i) ionizing
the Yb atoms into Yb$^{2+}$ + 2$e^-$  \cite{comm1}, (ii) condensing any excess
Yb
atoms into Yb metal, (iii) completely transfering the ionized electrons
from Yb to the C$_{60}$ molecules, and (iv) assembling the cations and
anions into a crystal. Since the electrostatic energy is the largest
single contribution to the structural energetics in these systems, it is
expected to dominate the energy differences between the configurations
of interest. Accordingly, we approximate the relative energies in steps (iii)
and
(iv) by the electrostatic energy differences alone. The ions are modeled
using point charges, and the Ewald method \cite{slater} is used to evaluate the
resulting Coulombic sums. Experimental structural parameters are used to
incorporate the effects of short-range repulsive interactions between
the Yb cations and C$_{60}$ anions. The parameters from Ref. 5 are slightly
simplified by taking {\em a} = {\em b} = {\em c} = 27.8733 \AA~(i.e., ignoring
the $<$
0.4\% orthorhombic distortion) and by assuming complete occupation of
the octahedral sites. All
energies are given in eV per Yb$_3$C$_{60}$ formula unit, or equivalently,
in eV/C$_{60}$.

We begin by computing the relative formation energies of  two
hypothetical  structures, ideal Yb$_3$C$_{60}$ and ``unrelaxed"
Yb$_{2.75}$C$_{60}$. The
latter is obtained by removing the requisite Yb
$T$-cations from an ideal Yb$_3$C$_{60}$ lattice to form the ordered- vacancy
structure  in  Fig \ref{fig1cap}(b), while keeping all  the remaining cations
at
the centers of their respective $O$- and $T$-site holes. The ionization steps
for ideal Yb$_3$C$_{60}$ and
unrelaxed Yb$_{2.75}$C$_{60}$ involve, respectively, 3 Yb atoms with an energy
$E_{\rm ion}$ = 55.29 eV and 2.75 Yb atoms with $E_{\rm  ion}$ = 50.68 eV
\cite{comm1,moore}. The
excess 0.25 Yb atom per C$_{60}$ in Yb$_{2.75}$C$_{60}$ is condensed into Yb
metal,
giving a cohesive energy gain $E_{\rm  coh}$ =  $-$0.4 eV \cite{kittel}. The
total
electron affinity for transferring 6 electrons to C$_{60}$ in Yb$_3$C$_{60}$ is
approximated as 6 times the first electron affinity for C$_{60}$ \cite{licht}
plus
the electronic self-energy, which is modeled as the minimum energy of 6
classical point electrons on a uniform spherical shell of radius 3.5 \AA
\cite{flem}; the result is a net energy cost $E_{\rm  ea}$ =  $-$15.90 + 41.04
= 25.14
eV. The analogous energy for transferring 5.5 electrons to C$_{60}$ in
Yb$_{2.75}$C$_{60}$ is approximated by interpolating the self-energy between
integer numbers of electrons $n$ = 2, 3, 4, and 6 with a second-order
polynomial in $n$, yielding $E_{\rm  ea}$ =  $-$14.58 + 33.17 = 18.59 eV.
Finally,
the Yb cations and C$_{60}$ anions are assembled into their respective
fcc-based crystal structures, differing only by the $T$-site vacancies in
unrelaxed Yb$_{2.75}$C$_{60}$. The Madelung energies, computed by treating the
Yb
cations as points with charge 2+ and replacing the spherical shells of
the C$_{60}$ anions with points of charge 6$-$  or 5.5$-$, are $E_{\rm  Mad}$ =
 $-$91.43 eV
for Yb$_3$C$_{60}$ and  $-$77.65 eV for Yb$_{2.75}$C$_{60}$.
Comparing the total formation energy $E_{\rm  form}$ = $E_{\rm  ion}$+$E_{\rm
coh}$+$E_{\rm  ea}$+$E_{\rm  Mad}$  =
$-$11.00 eV for ideal Yb$_3$C$_{60}$ with $E_{\rm  form}$ =   $-$8.78 eV for
unrelaxed
Yb$_{2.75}$C$_{60}$ shows that the larger Madelung energy gain
in forming Yb$_3$C$_{60}$ exceeds its larger ionization and self-energy costs,
and thus creating $T$-site vacancies in unrelaxed Yb$_{2.75}$C$_{60}$ is
endothermic:  $\Delta E_{\rm  form}$ = 2.22 eV.

The possible relaxation-energy gains
associated with vacancy formation must
now be considered. One obvious source is the displacement of each $O$-site
Yb cation towards its NN vacancy (see Fig \ref{fig1cap}(b)). The equilibrium
value
for this displacement, determined by the balance of electrostatic forces
and short-range repulsion between the cations and their neighboring
anions, is known from experiment to be extremely large (2.41 \AA
\cite{nature});
using this value leads to a computed energy gain of  $-$2.32 eV. The much
smaller off-center displacements of the $T$-site cations ($\sim$0.3 \AA)
contribute just  $-$0.009 eV on their own, but combined with those of the
$O$-site cations lead to a small cooperative interaction and a total
relaxation-energy gain $E_{\rm  displ}$ =  $-$2.40 eV.

The above calculations assumed that the charges on all C$_{60}$
anions are given by the average number of electrons per C$_{60}$, namely
$-$5.5$e$\cite{comm1}. 
The crystallographic inequivalence of the anions (see Fig
\ref{fig1cap}(c)),
however, makes favorable the prospect of transferring electrons from
C$_{60}${\em (1)}, with 12 NN cations, to C$_{60}${\em (3)}, which has only 8
\cite{citrin}. We
investigate this possibility by allowing the inequivalent C$_{60}${\em (1)},
C$_{60}${\em (2)}, and C$_{60}${\em (3)} anions to have correspondingly
different charges $-q_{\em 1}e$, $-q_{\em 2}e$, and $-q_{\em 3}e$, with the
only constraint being that ($q_{\em 1}$+6$q_{\em 2}$+$q_{\em 3}$)/8
= 5.5. Minimizing the quantity $E_{\rm  Mad}$ + $E_{\rm  ea}$ with respect to
$q_{\em i}$, we
find $q_{\em 1}$ =  6.3, $q_{\em 2}$ =  5.5 and $q_{\em 3}$ =  4.7.  The charge
of C$_{60}${\em (2)} is
unchanged, while 0.8$e^-$  is indeed transferred from C$_{60}${\em (3)} to
C$_{60}${\em (1)}. This
is not a charge disproportionation but is rather a charge
redistribution, reminiscent of the different charge states of the
crystallographically inequivalent planar and chain O anions in
YBa$_2$Cu$_3$O$_7$
\cite{citrin,comm2}. We calculate that in Yb$_{2.75}$C$_{60}$ the associated
energy gain $E_{\rm  redist}$ is just  $-$0.17 eV. This small value reflects
the nonlinear self-energy of electrons on C$_{60}$, and shows that the
redistribution process plays only a minor role in determining structural
stability.
The magnitude of the relaxation energy gain in Yb$_{2.75}$C$_{60}$ of $E_{\rm
displ}$ + $E_{\rm  redist}$ =  $-$2.57 eV is,
within the estimated accuracy of our calculations, essentially equal to
$\Delta$$E_{\rm  form}$ = 2.22 eV.

We now consider the additional relaxation energy that
can come from optimizing the orientation of the C$_{60}$ anions with respect
to their anisotropic environment of NN cations. The largest contribution
to the orientation energy is the electrostatic interaction between the
charge density of electrons transferred to C$_{60}$, anisotropically
distributed over its surface, and the neighboring Yb$^{2+}$ cations
\cite{wudl}. To
calculate this orientational interaction energy obviously
requires going beyond approximating the C$_{60}$ anions as uniformly charged
spherical shells. Based on comparisons with chemical analogs
\cite{citrin,wudl,taylor} and first-principles calculations of C$_{60}^{6-}$
\cite{and1}, which show that the
transferred charge density is primarily associated with the
pentagons, we model a C$_{60}$ anion as a truncated icosahedron with equal
point charges localized at the centers of its 12 pentagonal faces and
positioned 3.5 \AA~from its origin \cite{comm3}. For a single cation neighbor,
the C$_{60}$ anion will orient itself to minimize the distance between one of
the point charges and the cation, mimicking the preference for a cation
to be over a pentagonal face. Such an attraction between C$_{60}$ pentagonal
faces and neighboring cations was first observed in
(Ba$^{2+}$)$_3$C$_{60}^{6-}$ \cite{kortan} and
was found to be an important component in achieving the best Rietveld
refinement of the Yb$_{2.75}$C$_{60}$ structure \cite{nature,proc}.

The orientational interaction energy, $E_{\rm  \Theta}$, was investigated using
the set
of \{111\} axes established in the x-ray structural determination
\cite{nature}; these axes
are shown in Fig. \ref{fig1cap}(c). Each C$_{60}$ anion is rotated by the same
angle $\Theta$
around its own local axis.   $\Theta$ is taken to be zero in the uniform
reference configuration where the twofold axes of the C$_{60}$ anion are
aligned along the Cartesian axes \cite{comm4}. As shown in Fig. 2, $E_{\rm
\Theta}$
for both Yb$_3$C$_{60}$ and Yb$_{2.75}$C$_{60}$ depends strongly on $\Theta$
and exhibits a
pronounced minimum at  $\Theta$ = 37.5$^\circ$, consistent with the x-ray
refinement
results \cite{nature,proc}. However, because of the large displacements of the
$O$-site cation in Yb$_{2.75}$C$_{60}$, a greater degree of pentagon-cation
interaction is achieved through this orientation of the C$_{60}$ anions.
Relative to ideal Yb$_3$C$_{60}$, the orientational relaxation-energy gain
 in
Yb$_{2.75}$C$_{60}$ is appreciable:  $\Delta$$E_{\rm  \Theta}$ =  $-$0.97 eV.
While we expect our simple point-charge model for distributing charge on
C$_{60}$ to overestimate the absolute magnitude of the orientational
dependence of the energy, the scale of the calculated dif-
\begin{figure}
\epsfxsize=3.2 truein
\centerline{\epsfbox{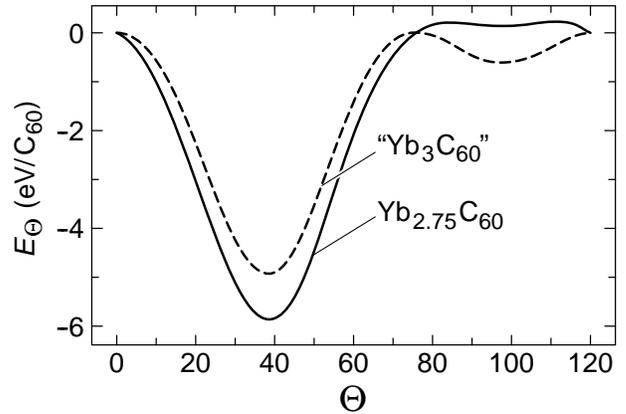}}
\caption{Angular dependence of orientational interaction energy in the
Yb$_{2.75}$C$_{60}$ and ideal Yb$_3$C$_{60}$ structures due to rigid rotation
of all C$_{60}$ anions about local \{111\} axes.}
\label{fig2}
\end{figure}
\noindent
ference $\Delta$$E_{\rm  \Theta}$ demonstrates the importance of this effect
for stabilizing the observed vacancy-ordered phase.

A final source of relaxation energy can arise from distorting the
C$_{60}$ anions. The anisotropic crystal field of NN Yb cations (see
Fig.\ref{fig1cap}(c)) is expected to distort the symmetric shape of a C$_{60}$
anion even with a
uniform charge distribution, and such distortions will be still greater
when the charge distribution is itself anisotropic
\cite{and1,citrin,wudl,taylor}. Indeed, recent x-ray absorption and Raman
measurements in Yb$_{2.75}$C$_{60}$
show clear evidence for significant
shape deformations of the C$_{60}$ anions
\cite{citrin}. We calculate the relative distortion-energy gain $\Delta E_{\rm
dis}$ using the same point-charge model as in the calculation of $E_{\rm
\Theta}$, with the charges allowed to move radially relative to their ideal
positions in the $\Theta$ = 37.5$^\circ$ orientation. The stiffness of the
C$_{60}$ anion is modeled by connecting the charges to the anion center with
harmonic
springs whose spring constant is chosen to reproduce the observed $A_g(1)$
radial breathing mode frequency of $\sim$500 cm$^{-1}$ \cite{rev2}.
The associated relaxation-energy gain is $-$0.32 eV in Yb$_{2.75}$C$_{60}$ and
$-$0.28 eV in
Yb$_3$C$_{60}$, giving $\Delta E_{\rm dis}$ = 0.04 eV.
In both systems, outward displacements range from 0.01 \AA~to 0.04 \AA.
While these displacements are somewhat smaller than those inferred  from the
EXAFS data \cite{citrin}, suggesting that the absolute distortion energies are
underestimated by our model, the scale of the difference $\Delta E_{\rm dis}$
indicates that the distortion contribution to the total relaxation energy is
relatively minor.

Our comparison of relevant energies in forming the Yb$_{2.75}$C$_{60}$ and
Yb$_3$C$_{60}$ structures provides strong support for the central role played
by
the interaction between the anisotropically positioned Yb cations and
the anisotropically distributed charge on the C$_{60}$ anions. As shown here,
this strong interaction can be investigated through a relatively
straightforward analysis of electrostatic energies, and does not require
invoking any chemical behavior peculiar to Yb. Despite its simple form,
the interaction has a profound effect on compound formation and crystal
structure. In particular, application of our model gives insight into
the stability of a seemingly unwieldy structure involving
both long-range-ordered vacancies
in a 2008-atom unit cell and crystallographically inequivalent C$_{60}$
anions with low-symmetry cation coordinations. The relationship between
these unusual structural features is now clear: A $T$-site cation vacancy
leads to the $O$-site cation displacements, which in turn amplify the
orientational cation-anion interactions.
Only one vacancy
per subcell is needed to induce displacements of {\em all} $O$-cations, making
the formation of additional vacancies unfavorable.  This naturally accounts for
the observed Yb$_{2.75}$C$_{60}$ stoichiometry.

One consequence of this simple picture is that the same
ordered-vacancy phase for Yb$_{2.75}$C$_{60}$ should be observed in fulleride
compounds containing other divalent cations whose
ionic radii and elemental cohesive energies are similar to Yb. Such
expectations are, in fact, realized for both Ca- \cite{proc,ca} and Sm-doped
\cite{chen} C$_{60}$. Since the magnitudes of the electrostatic energies scale
quadratically with charge, it might also be expected that many of the
stabilizing factors identified here for these divalent-cation fullerides
should occur for the similarly-sized monovalent Na cation. However,
neither a Na$_3$C$_{60}$ structure nor a vacancy-ordered fulleride phase is
observed; instead, a structure in which two Na cations occupy the same
$O$-site hole is found \cite{ross}. The reason for this is that other factors,
such as the short-range repulsive interactions between cations and C$_{60}$
anions and between cations in the same $O$-site \cite{and2}, scale differently
with charge and affect the structural stability of this system.

The question of off-center displacements for $O$-site cations has
recently been a subject of debate in several fcc-based alkali fullerides
\cite{now,bend}. For those systems in which $T$-site vacancies are observed,
e.g.,
Rb$_3$C$_{60}$ \cite{fischer}, our analysis leads unavoidably to the conclusion
that the
surrounding $O$-site cations will undergo off-center displacements towards
that vacancy (see Fig \ref{fig1cap}(b)). While this applies only to the
fraction of
$O$-site cations immediately surrounding the vacancies, such displacements
should be detectable with appropriate measurements.

The model presented here is designed to capture the essential  features of the
structural energetics of the systems of interest, and to enhance
the understanding of the origin and properties of complex
structures. In Yb$_{2.75}$C$_{60}$, we found that the relaxation energies
associated with the lowered symmetry are crucial in stabilizing its
formation. The quantitative limitations of this type of analysis are
clear. However, our model can provide valuable guidance for future, more
definitive investigations of structural energetics in this and similar
systems using first-principles
calculations, and for designing and
interpreting related experiments in the interim.

We thank E. \"Ozda\c{s}, R. C. Haddon, W.
Andreoni, S. C. Erwin, E. J. Mele and P. B. Littlewood for helpful discussions.
K. R. acknowledges the Aspen Center for Physics and the Alfred P. Sloan
Foundation.


\begin{references}
\bibitem{comm1}  Yb$^{3+}$ ions are not observed in Yb$_{2.75}$C$_{60}$ (cf.
Refs \onlinecite{nature} and \onlinecite{citrin}), and are assumed absent in
Yb$_3$C$_{60}$.
\bibitem{comm3} Charge redistribution in Yb$_{2.75}$C$_{60}$ is ignored here.
\bibitem{comm4} The two orientations consistent with this specification are
symmetry-equivalent in our simplified cubic structure.
\end{references}
\end{document}